\documentclass[conference]{IEEEtran}

\usepackage{algorithmic}
\usepackage{amsmath,amssymb,amsfonts}
\usepackage{balance}
\usepackage{cite}
\usepackage{comment}
\usepackage{dirtree}
\usepackage{fontawesome}
\usepackage[T1]{fontenc}
\usepackage{graphicx}
\usepackage{hyperref}
\usepackage[linewidth=1pt]{mdframed}
\usepackage{multirow}
\usepackage{pgfplots}
\usepackage{pifont}
\usepackage{setspace}
\usepackage{subcaption}
\usepackage{tablefootnote}
\usepackage{textcomp}
\usepackage{tikz}
\usetikzlibrary{patterns}
\pgfplotsset{compat=1.18}
\usepackage{url} 
\usepackage{xcolor}
\usepackage{xspace}

\usepackage[acronym]{glossaries}
\newacronym{ann}{ANN}{Approximate Nearest Neighbor}
\newacronym{asr}{ASR}{Automatic Speech Recognition}
\newacronym{ir}{IR}{Information Retrieval}
\newacronym{nlp}{NLP}{Natural Language Processing}
\newacronym{rag}{RAG}{Retrieval Augmented Generation}
\newacronym[plural=LLMs,firstplural=Large Language Models(LLMs)]{llm}{LLM}{Large Language Model}
\newacronym{ui}{UI}{user interface}
\newacronym{cli}{CLI}{command-line interface}

\def\BibTeX{{\rm B\kern-.05em{\sc i\kern-.025em b}\kern-.08em
    T\kern-.1667em\lower.7ex\hbox{E}\kern-.125emX}}

\definecolor{forestgreen}{RGB}{34,139,34}
\definecolor{darkyellow}{RGB}{255,223,0}
\definecolor{debianred}{rgb}{0.84, 0.04, 0.33}
\definecolor{midnightblue}{HTML}{006795}
\definecolor{royalpurple}{HTML}{613F99}

\newcommand{\frameworkname}[0]{SoccerRAG\xspace}

\newcommand{\gptfour}[0]{GPT-4.0-Turbo\xspace}
\newcommand{\gptthree}[0]{GPT-3.5-Turbo\xspace}

\newcommand{\figw}{.99\columnwidth}

\begin{document}
\bstctlcite{IEEEexample:BSTcontrol}

\title{Demo: Soccer Information
Retrieval via Natural Queries using \frameworkname}

\author{\IEEEauthorblockN{Aleksander Theo Strand}
\IEEEauthorblockA{\textit{OsloMet, TET Digital AS}\\
Oslo, Norway \\
0009-0008-2749-2347}
\and
\IEEEauthorblockN{Sushant Gautam}
\IEEEauthorblockA{\textit{OsloMet, SimulaMet} \\
Oslo, Norway \\
0000-0001-9232-2661}
\and
\IEEEauthorblockN{Cise Midoglu}
\IEEEauthorblockA{\textit{SimulaMet, Forzasys} \\
Oslo, Norway \\
0000-0003-0991-4418}
\and
\IEEEauthorblockN{Pål Halvorsen}
\IEEEauthorblockA{\textit{OsloMet, SimulaMet, Forzasys} \\
Oslo, Norway \\
0000-0003-2073-7029}
}

\maketitle

\begin{abstract}
The rapid evolution of digital sports media necessitates sophisticated information retrieval systems that can efficiently parse extensive multimodal datasets. This paper demonstrates \frameworkname, an innovative framework designed to harness the power of \gls{rag} and \glspl{llm} to extract soccer-related information through natural language queries. By leveraging a multimodal dataset, \frameworkname supports dynamic querying and automatic data validation, enhancing user interaction and accessibility to sports archives. We present a novel interactive \gls{ui} based on the Chainlit framework which wraps around the core functionality, and enable users to interact with the \frameworkname framework in a chatbot-like visual manner.
\end{abstract}


\glsresetall
\section{Introduction}\label{section:introduction}

The burgeoning interest in \gls{rag}, fueled by the rapid advancements in \glspl{llm}, has paved new avenues for exploring innovative use cases across various domains, notably within multimodal information retrieval~\cite{Cheng2024Jan}. Despite the wide applicability of \gls{rag} frameworks in enhancing the capabilities of generative AI for open-domain question answering and beyond, its potential in the sports domain, particularly within soccer analytics, remains largely untapped. Our recent research has endeavored to bridge this gap by harnessing multimodal datasets inherent to the soccer broadcast pipeline and integrating \gls{rag} with different data modalities. We have presented \frameworkname~\cite{SoccerRAG-CBMI}, a framework for retrieving multimodal soccer information using natural language queries, from an augmented soccer dataset based on SoccerNet~\cite{SoccerNet,SoccerNetv2}, which includes game videos with image frames and audio, timestamped captions (transcribed audio), annotations for game events, and player information. 

\begin{figure}
    \centering
    \includegraphics[width=\columnwidth]{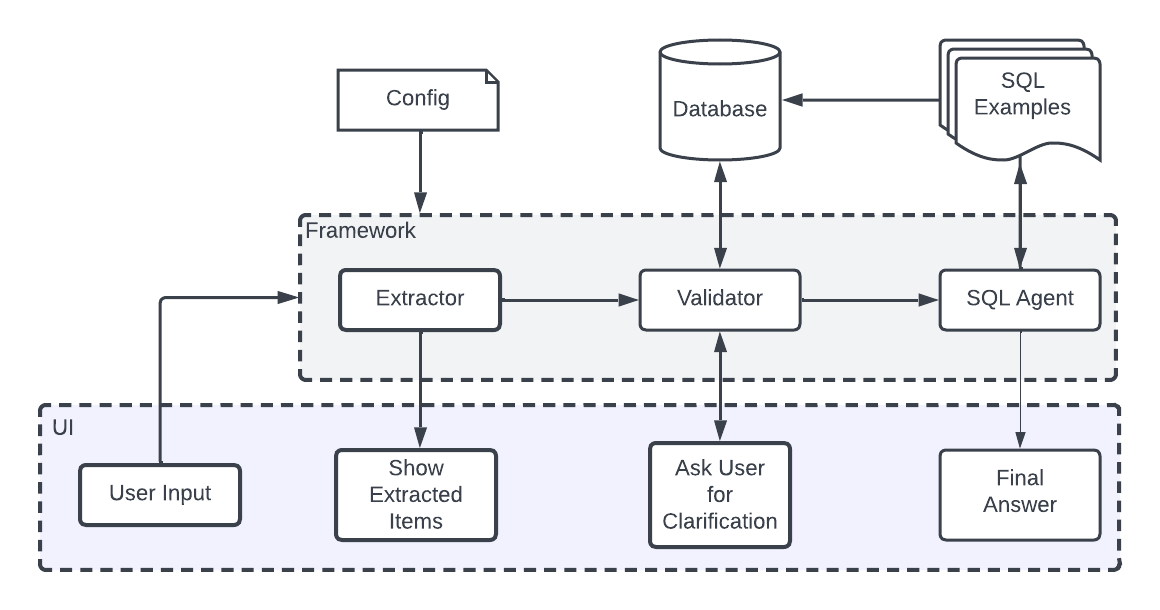}
    \caption{Overview of the \frameworkname framework, including the novel interactive \gls{ui}.}
    \label{fig:demo-framework}
\end{figure}

In this paper, we demonstrate the use of the \frameworkname framework through a \gls{ui} based on ChainLit~\cite{ChainlitOverview}, as well as the \gls{cli}. Figure~\ref{fig:demo-framework} presents an overview of the \frameworkname framework, with the core components (database, feature extractor, feature validator, and SQL agent) as presented earlier, wrapped by a novel interactive \gls{ui} which serves as a user-friendly tool for visual interactions. The \gls{ui} was created using the Chainlit framework. Chainlit is an open-source Python package that enables developers to build production-ready conversational AI applications, providing features for quick integration. It offers integrations with popular libraries and frameworks, such as OpenAI and LangChain, and allows for custom front-ends with React-based user interfaces~\cite{ChainlitOverview}. The \frameworkname integrates functionality for data representation, feature extraction and validation, database querying (based on our proposed database schema for the augmented version of the SoccerNet dataset), and a novel extractor-validator chain. The open source implementation for \frameworkname is accessible under~\cite{SoccerRAG-GitHub} and fully reproducible through the instructions provided therein, which are detailed below.

The potential of our proposed system for information retrieval offers a promising avenue for fans and broadcasters alike, enabling the recollection of game specific information and highlights, as well as broader statistics and insights (within the scope of the entire database) through natural queries. Pursuing this avenue of research, we aspire not only to contribute to the body of knowledge surrounding \gls{rag} and multimodal information retrieval, but also to pioneer a smart system that encapsulates the dynamic essence of soccer, fostering a deeper connection between the sport and its global audience. 

\section{Demonstration}\label{section:demonstration}

\subsection{Application Flow}\label{sec:application-flow}

The core functionality of the \frameworkname framework can be represented with the following application flow:

\begin{itemize}
    \item The user provides a natural language query related to the contents of the database. 
    \item The user input is sent to an \gls{llm} along with the properties schema and a system prompt describing the properties the \gls{llm} should extract from the query. The \gls{llm} then returns a list of extracted properties relevant to the query. 
    \item Each extracted feature is checked against the appropriate table in the database using string matching algorithms. This step aims to correct spelling mistakes and abbreviations. Once a value is found, both the value and its primary key are added to the extracted value. 
    \item The cleaned user prompt is combined with system-specific prompts to guide the \gls{llm} in generating SQL queries that will answer the user's natural language query. The constructed query is then passed to the SQL chain, which designs and executes the SQL queries against the underlying database. The SQL chain handles communication between the system and the database, retrieves the requested data, and prepares the results for presentation to the user. 
\end{itemize}

\subsection{Artifacts}

The \frameworkname codebase is publicly accessible under~\cite{SoccerRAG-GitHub}. The \texttt{data} folder is used by the core pipeline to read in the source dataset (including the original SoccerNet data, as well as our augmented league and team tables), and storing the database files. The \texttt{src} folder contains the configuration and executable files. The root directory includes the main executable (\texttt{main.py}), environment variables (\texttt{.env}), and a list of dependencies (\texttt{requirements.txt}). A proof-of-concept deployment of the \frameworkname \gls{ui} can be found under: \url{https://simulamet-host-soccerrag.hf.space/}

\subsection{Requirements}

The \frameworkname codebase requires Python 3.12 or above to be run. The \texttt{requirements.txt} file in the codebase specifies the required packages, which can be installed through \texttt{pip}. \frameworkname does not require a GPU, and can be run on any machine with a CPU. For the experiments presented in~\cite{SoccerRAG-CBMI}, a machine with a Windows 11 operating system, Intel(R) Core(TM) i5-9300H CPU @2.40GHz, and 16GB memory has been used. \frameworkname requires an OpenAI API key, which should be specified in the \texttt{.env} file. In order to log all queries to the OpenAI endpoints, it is possible to set a LangSmith API key~\cite{LangSmith}. This makes it possible to monitor the cost of each call, and keep history.

\subsection{Running the Framework}

Below are the steps to run the \frameworkname framework (core functionality described in Section~\ref{sec:application-flow}).

\subsubsection{Configuration} 

The \frameworkname framework is set up using environment variables. Table~\ref{tab:config-parameters} presents an overview of the configuration parameters. The parameters marked as mandatory do not have default values, and need to be configured before runtime. There should be an entry in \texttt{.env} for each configuration parameter. The \frameworkname framework has been tested with \gptthree~\cite{OpenAIGPT3} and \gptfour~\cite{OpenAIGPT4,OpenAIGPT4Docs2023}, but can support any model available from OpenAI~\cite{OpenAIModels}. The model version can be specified through the \texttt{OPENAI\_MODEL} configuration parameter. 

\begin{table}[htbp]

    \scriptsize
    \centering
    
    \begin{tabular}{@{\hspace{0mm}}l@{\hspace{2mm}}l@{\hspace{2mm}}c@{\hspace{2mm}}l@{\hspace{0mm}}}
        \hline
        
        \textbf{Parameter} 
        & \textbf{Description} 
        & \textbf{M}
        & \textbf{Default} 
        \\ \hline
        
        \texttt{OPENAI\_API\_KEY} 
        & OpenAI API key 
        & \ding{52}
        & NA 
        \\ \hline
        
        \texttt{OPENAI\_MODEL} 
        & GPT version 
        & \ding{54}
        & gpt-3.5-turbo-0125 
        \\ \hline
        
        \texttt{DATABASE\_URL} 
        & Location of DB 
        & \ding{54}
        & "/data/games." 
        \\ \hline
        
        \texttt{LANGSMITH}
        & Trace api calls 
        & \ding{54} 
        & False 
        \\ \hline
        
        \texttt{LANGSMITH\_API\_KEY} 
        & API key for LangSmith 
        & \ding{52}\tablefootnote{If \texttt{LANGSMITH} is set to True.}
        & NA 
        \\ \hline
        
        \texttt{LANGSMITH\_PROJECT} 
        & Project tag for LangSmith 
        & \ding{54}
        & SoccerRag
        \\ \hline

        \texttt{FEW\_SHOT} 
        & Num. options in user validation  
        & \ding{54}
        & 3
        \\ \hline

    \end{tabular}
    
    \caption{\frameworkname parameters (M: mandatory).}
    \label{tab:config-parameters}
    
\end{table}

\subsubsection{Running via \gls{ui}}

Run \texttt{python setup.py} to download the dataset and set up the DB, and \texttt{chainlit run app.py} for the \gls{ui}.

\begin{figure}
    \scriptsize
    \centering
    \frame{\includegraphics[width=\figw]{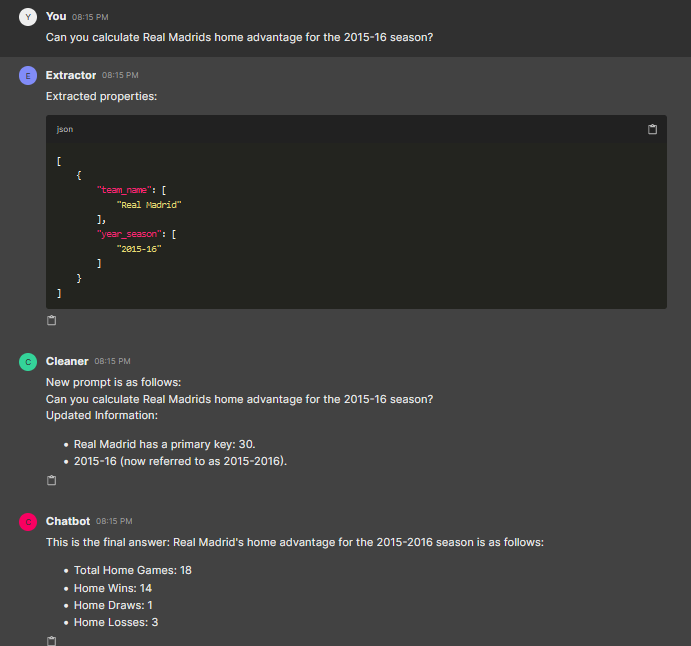}}
    \caption{\frameworkname \gls{ui} example 1. The user prompts: "Can you calculate Real Madrids \textit{[sic]} home advantage for the 2015/16 season?"}
    \label{fig:gui-1}
\end{figure}

Figure~\ref{fig:gui-1} presents an example screenshot\footnote{\frameworkname \gls{ui} background color is configurable: click the user icon in the top right corner, select settings, then toggle the dark mode switch.} depicting the intermediate feedback to users that the \frameworkname provides for each operational step, which are retrieved from the feature extractor and feature validator components in the core pipeline. The meaningful response returned to the user prompt "Can you calculate Real Madrids \textit{[sic]}\footnote{\textit{sic} (adverb) indicates that the quoted matter is exactly as found in the original source, including erroneous spelling, punctuation, or grammar.} home advantage for the 2015/16 season?" demonstrates that the framework is able to grasp soccer-related concepts such as a home advantage (which refers to the difference in scores achieved by a team in home vs. away games). 

\begin{figure}
    \scriptsize
    \centering
    \frame{\includegraphics[width=\figw]{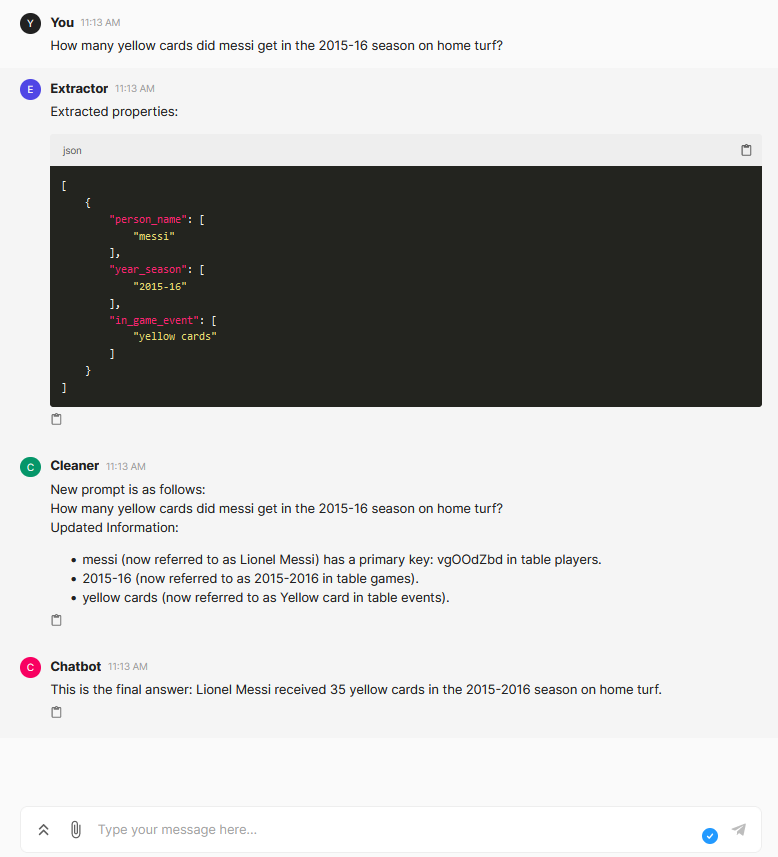}}
    \caption{\frameworkname \gls{ui} example 2. The user prompts: "How many yellow cards did messi \textit{[sic]} get in the 2015-16 season on home turf?"}
    \label{fig:gui-2}
\end{figure}

\begin{figure}
    \scriptsize
    \centering
    \frame{\includegraphics[width=\figw]{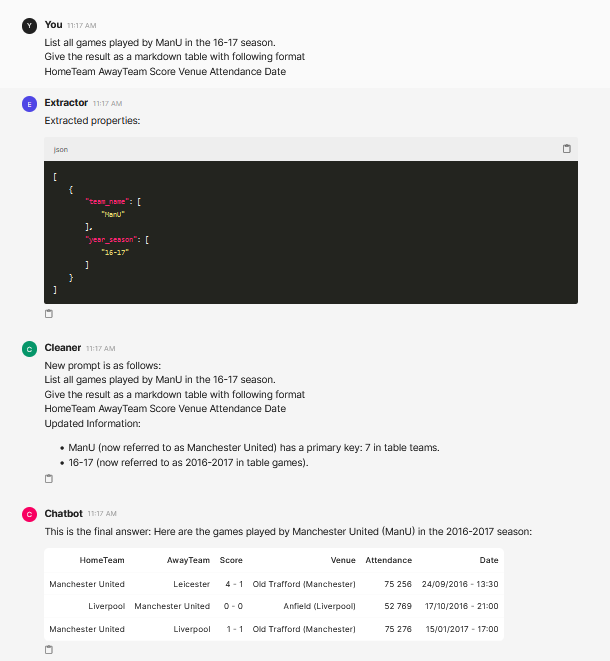}}
    \caption{\frameworkname \gls{ui} example 3. The user prompts: "List all games played by ManU \textit{[sic]} in the 16-17 season. / Give the result as a markdown table with following format / HomeTeam AwayTeam Score Venue Attendance Date"}
    \label{fig:gui-3}
\end{figure}

Figures~\ref{fig:gui-2} and \ref{fig:gui-3} present further example screenshots, demonstrating the capabilities of \frameworkname which range from database-wide statistics (example 2: "How many yellow cards did messi \textit{[sic]} get in the 2015-16 season on home turf?" is returned with the correct aggregation across all games of the desired season), to supporting multi-format information presentation (example 3: "List all games played by ManU \textit{[sic]} in the 16-17 season. / Give the result as a markdown table with following format / HomeTeam AwayTeam Score Venue Attendance Date"). 

\begin{figure}
    \centering
    \begin{subfigure}[b]{\figw}
        \centering
        \frame{\includegraphics[width=\columnwidth]{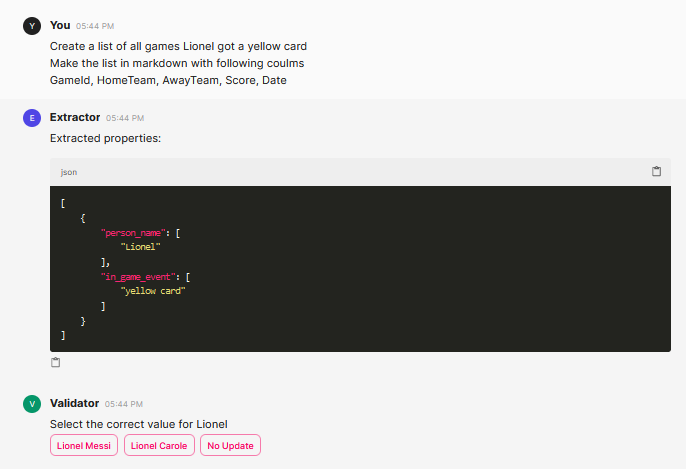}}
        \caption{Step 1: user asked for clarification}
        \label{fig:gui-4-step1}
    \end{subfigure}
    \begin{subfigure}[b]{\figw}
    \centering
        \frame{\includegraphics[width=\columnwidth]{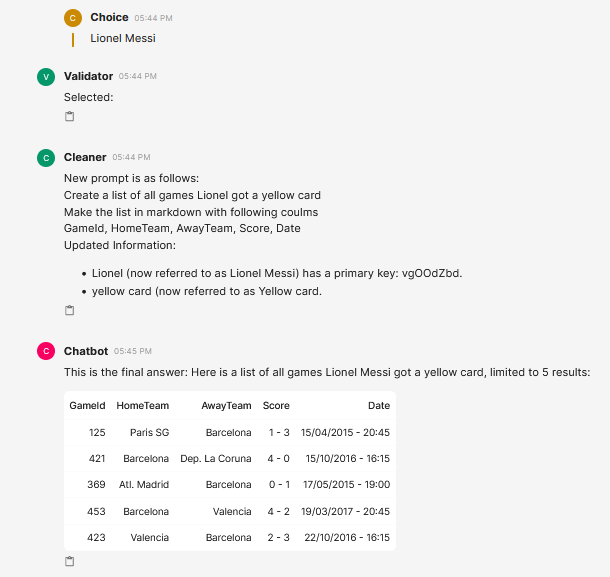}}
        \caption{Step 2: response returned based on additional user input}
        \label{fig:gui-4-step2}
    \end{subfigure}
    \caption{\frameworkname \gls{ui} example 4. The user prompts: "Create a list of all games Lionel got a yellow card / Make the list in markdown with following coulms \textit{[sic]} / GameId, HomeTeam, AwayTeam, Score, Date".}
    \label{fig:gui-4}
\end{figure}

One of the challenges faced by systems which are prompted via natural language queries is the potential lack of clarification with respect to the entities in the prompt. In the context of soccer, these can be player, team, or league names, places, season dates, etc. \frameworkname requests clarification from the user in cases of inadequate information, as shown in Figure~\ref{fig:gui-4}. Here (example 4: "Create a list of all games Lionel got a yellow card / Make the list in markdown with following coulms [sic] / GameId, HomeTeam, AwayTeam, Score, Date"), as opposed to directly returning a response ("one-step") as in examples 1-3, the framework needs to ask the user to select the correct value for "Lionel", as there are multiple entities in the database with this name ("two-step"). In the \gls{ui}, the user can click a button to communicate their choice and validate one of the presented options, or pass the original string untranslated to the next component.

\subsubsection{Running via \gls{cli}}

Run \texttt{python setup.py} to download the dataset set up the DB, and \texttt{python main\_cli.py -q <query>} to use the \gls{cli}. The \texttt{<query>} should be encapsulated by quotation marks (e.g., "How many goals did Arsenal score in the 2015-16 season?"). If linebreaks are intended, \texttt{"\textbackslash n"} can be used.

\begin{figure}
    \scriptsize
    \centering
    \frame{\includegraphics[width=\figw]{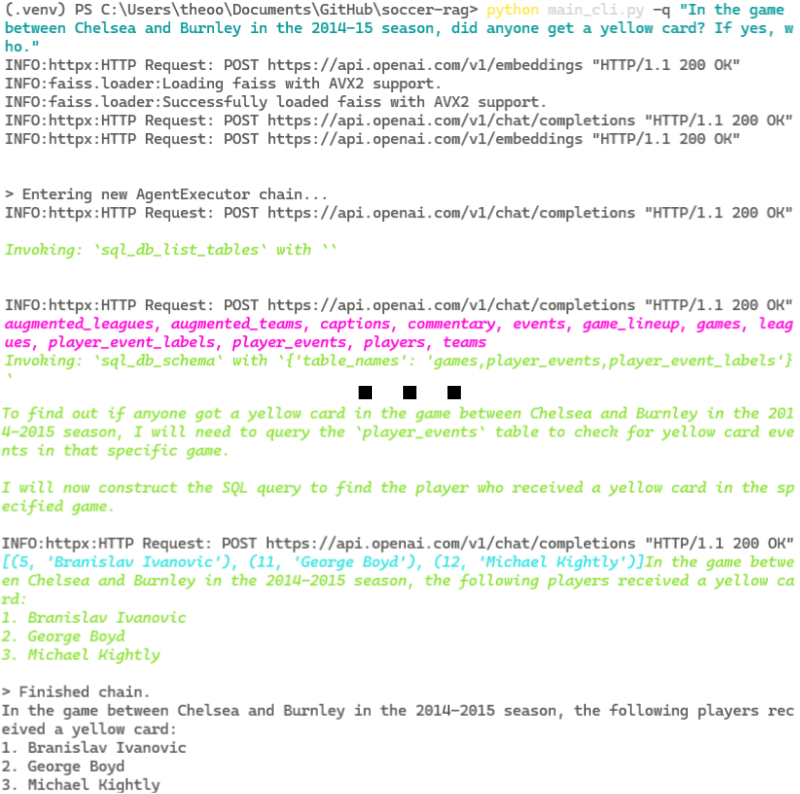}}
    \caption{\frameworkname \gls{cli} example 1. The user prompts: "In the game between Chelsea and Burnley in the 2014-15 season, did anyone get a yellow card? If yes, who."}
    \label{fig:cli-1}
\end{figure}

Figure~\ref{fig:cli-1} presents an example screenshot, depicting the console outputs from the \gls{cli} for the query "In the game between Chelsea and Burnley in the 2014-15 season, did anyone get a yellow card? If yes, who.", which receives a response directly (one-step). 

\begin{figure}
    \scriptsize
    \centering
    \frame{\includegraphics[width=\figw]{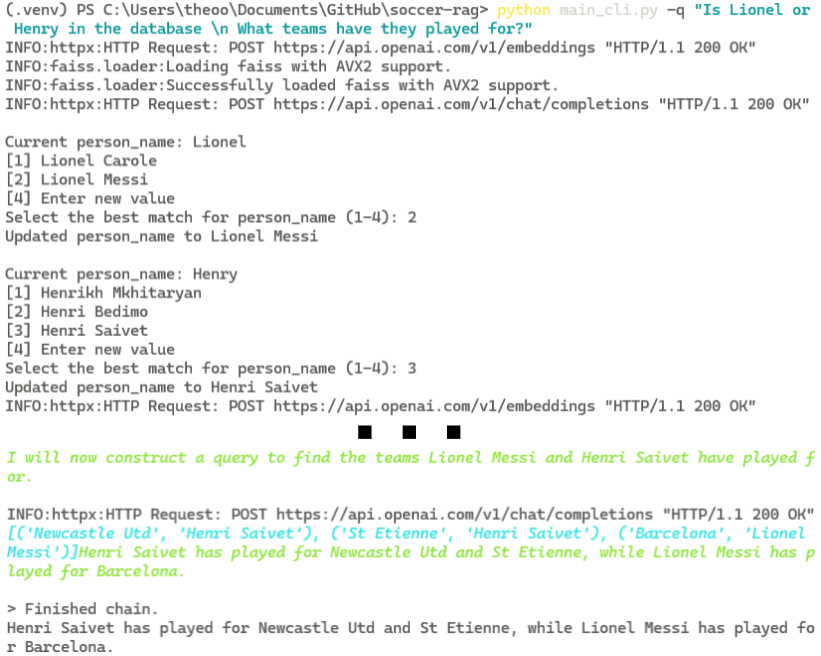}}
    \caption{\frameworkname \gls{cli} example 2. The user prompts: "Is Lionel or Henry in the database \textbackslash n What teams have they played for?"}
    \label{fig:cli-2}
\end{figure}

Figure~\ref{fig:cli-2} presents an example screenshot, depicting the console outputs for the query "Is Lionel or Henry in the database \textbackslash n What teams have they played for?", which requires clarification from the user before returning the response (two-step). Differently than the \gls{ui}, the user can also provide a custom string as a new value (option). 

\subsubsection{Results}

In~\cite{SoccerRAG-CBMI}, we demonstrate that this framework can improve the hit-rate of the answers from 20\% to 80\%.

\section{Conclusion}\label{section:conclusion}

We demonstrate \frameworkname, a framework that leverages \gls{rag} and \glspl{llm} to efficiently retrieve multimodal soccer information via natural language queries, enhancing the accessibility of sports datasets by allowing for intuitive user interactions with complex data archives. In addition to the core pipeline presented in~\cite{SoccerRAG-CBMI}, we introduce a novel \gls{ui} to facilitate visual user interactions with the framework in a chatbot-like manner. The \gls{ui} serves as a wrapper around the core pipeline, whereby it passes the user query to the feature extractor component, after which the query is validated by the feature validator against the database, the query with validated properties are passed to SQL agent, which retrieves data from the database and returns a response, and the \gls{ui} presents the response to the user. Our experiments up to now demonstrate \frameworkname's capability to accurately interpret complex queries and facilitate dynamic user engagements with soccer content. 

We present the entire codebase for the \frameworkname as open source software, along with instructions to run the framework either via \gls{cli} or via \gls{ui}. The underlying framework concept can be used for other applications as well, provided that a source dataset and SQL schema are available. Looking forward, \frameworkname is poised for enhancements such as real-time processing and integration with emerging AI technologies, which will broaden its application beyond soccer, or even sports, and improve user experience. 

\section*{Acknowledgment}

This research was partly funded by the Research Council of Norway, project number 346671 (AI-Storyteller). 

\clearpage
\balance
\bibliographystyle{IEEEtran}
\bibliography{references}

\end{document}